\documentclass[twocolumn,floatfix]{revtex4}
\usepackage{epsfig}
\usepackage{graphicx}
\usepackage{dcolumn}
\usepackage{amsthm,amsmath}

\begin{document}

\title{A Relativistic Equation-of-motion Coupled-Cluster Investigation of the Trends of Single and Double Ionization 
Potentials in the He and Be Isoelectronic Systems} 

\author{Himadri Pathak\footnote{Email: h.pathak@ncl.res.in}$^1$,  B. K. Sahoo\footnote{Corresponding email: bijaya@prl.res.in}$^2$,  Turbasu Sengupta$^1$, B. P. Das$^3$, Nayana Vaval$^1$ and Sourav Pal$^1$}

\affiliation{$^1$Electronic Structure Theory Group, Physical Chemistry Division, CSIR-National Chemical Laboratory, Pune, 411008, India}

\affiliation{$^2$Theoretical Physics Division, Physical Research Laboratory, Ahmedabad-380009, India}

\affiliation{$^3$Theoretical Physics and Astrophysics Group, Indian Institute of Astrophysics, Bangalore-560034, India}

\begin{abstract}
We employ four-component spinor relativistic equation-of-motion coupled-cluster (EOMCC) method within the single- and double- 
excitation approximation to calculate the single ionization 
potentials (IPs) and double ionization potentials (DIPs) of the He and Be isoelectronic sequences up to Ne.
The obtained results are compared with the available results from the National Institute of Standards and
Technology (NIST) database to test the performance of the EOMCC method.
We also present intermediate results at different level of approximations in the EOMCC framework to gain
insight of the effect of electron correlation. 
Furthermore, we investigate the dependence of the IPs and DIPs of these ions on the ionic charge and observe that these
follow parabolic trends. Comparison between the trends of IPs and DIPs in both the classes of considered systems
are categorically demonstrated. 
\end{abstract}

\maketitle
\section{Introduction}
There have been significant advances in the experimental techniques in the recent years like time-of-flight 
photoelectron-photoelectron coincidence (TOF-PEPECO) spectroscopy \cite{1,2} which enables to carry out 
high precision measurements of multi-ionization processes stimulated by single photons from synchrotron 
sources. Sources like x-ray free electron laser of Linac Coherent Light sources of SLAC \cite{3,4} are capable of 
producing sequential and direct multiphoton-multielectron interactions.
A single photon having sufficient energy may 
eject two electrons to the continuum orbitals which can be treated as a topological three-body quantal problem.
The mutual interaction between the electrons (electron correlation) play a major role
in these processes.

The multiply charged ions are of significant interest in diverse areas of physics, starting from the x-ray space 
astronomy, plasma physics to laser physics \cite{7,8}. In a recent work \cite{9}, it has been demonstrated that 
the highly charged ions of C, N, O and Ne are the constituents of warm-hot intergalactic medium and are distributed 
mainly between their He and H like species. These highly charged ions have significance in determining the mass of 
missing baryons \cite{10,11}. 

The wave function constructed using a non-relativistic theory is not adequate for the accurate description of the energy 
spectrum of multiply charged ions, and a relativistic description is imperative in this case \cite{12}. On the other hand, 
accurate description of the relaxation effects along with the electron correlation effects are essential for explaining 
multiple ionization processes precisely. Therefore, calculations based on the lower order many-body theories are not 
reliable enough and may not be able to complement to describe the quality of results that are anticipated from the 
on-going sophisticated experiments \cite{13}. Kaldor and coworkers have made significant contributions in 
addressing the simultaneous treatment of relativistic and electron correlation effects using 
coupled-cluster (CC) methods.
They extended the effective Hamiltonian variant of the Fock-space multi-reference theory to the relativistic domain and
applied to both atomic and molecular systems \cite{14,15,16,17,18,19}. A brief overview of the effective 
Hamiltonian variant of FSMRCC is given in \cite{45}. A detail description of the effective Hamiltonian variant 
of the FSMRCC can be found in Refs. \cite{20,21,22,23,24,25,26,27}. The challenging problem associated with the
effective Hamiltonian theory is the problem of intruder states which arises due to the presence of 
quasi-degeneracy in the eigenstates lead to the failure in convergence \cite{28,29,30,31,32}. The intermediate 
Hamiltonian variant of the FSMRCC (IHFSMRCC) theory is a solution to address the problem of intruder states 
\cite{33,34,35,36,37,38,39,40}. The eigenvalue independent partitioning technique (EIP) of Mukherjee in the 
multi-reference coupled-cluster (EIP-MRCC) approach is also a specific variant of IHFSMRCC to take care of the 
intruder state problem \cite{41,41a}. It converts the non-linear FSMRCC equations for a given model space into 
a set of linear CI-like non hermitian eigenvalue equations. The solution can be reached by solving one root at 
a time preserving the norm of the corresponding eigenvector. This makes EIP-MRCC free from the problem of 
intruder and achieving faster convergence as compared to the effective Hamiltonian based FSMRCC. 

On the other hand fully four-component relativistic method 
in the equation-of-motion coupled-cluster (EOMCC) framework \cite{42,43,44,45} is an alternative potential choice to 
treat different many-body effects in a balanced manner.
It describes the complex multi-configurational wave function within a single reference description.
The key feature of the EOMCC method is that its
reference wave function is obtained using the CC method, which takes care of the dynamical parts of the electron correlations while the non-dynamic counterparts
are incorporated through the diagonalization of matrix elements of an effective Hamiltonian in the configuration 
space \cite{46,47,48,49,50,51,52,52a}. Furthermore, the EOMCC method is free from the intruder state problem 
due to CI like structure. This method scales properly 
at the non-interacting limit and does not satisfy the requirement of linkedness rigorously ensuring the size extensivity 
\cite{53,53e}.
It is worth to mention that EOMCC method is equivalent to effective Hamiltonian variant of the FSMRCC for the 
single ionization or attachment problem \cite{53d}, whereas it is not so in the case of double ionization or attachment (DI/DE-EOMCC) \cite{53a,53b,53c}. 
The reason behind is that the FSMRCC theory requires the amplitude equations of all
the lower sectors along with the amplitudes of the 
sector of interest, whereas EOMCC requires only the amplitudes of the (0,0) sector and that of the considered sector.

Another important aspect of EOMCC is, it directly gives the eigenstates in contrast to the propagator based
approaches \cite{54,55} although these also have equation-of-motion (EOM) structure.
Coupled cluster linear response theory (CCLRT) \cite{56,57,58,59,59a} is closely related to EOMCC in nature and produces identical results for 
the energy differences. A comparative discussion on EOMCC and CCLRT can be found in \cite{60,61,62}.
Choudhury {\it et al,} applied the relativistic CCLRT to calculate ionization potentials and related phenomena \cite{63,64,65,66,67,67a}.
Symmetry adapted cluster expansion configuration interaction (SAC-CI) \cite {68,69,70} is also very similar to EOMCC. These
two methods differ in the description of the ground state wave-function.
 
In this work, we employ our recently developed fully four-component relativistic EOMCC method to calculate core as 
well as valence ionization potentials (IPs) and double ionization potentials (DIPs) of He-like and Be-like atomic systems with 
atomic number $Z \leq 10$. 
The results of our calculations are compared with the available results from the National Institute of Standards and Technology 
(NIST) database \cite{71}. To understand the role of electron correlations, we followed different approximate schemes in the EOMCC framework. 
We call these as MBPT2-RPA, MBPT2-EOMCC and CCSD-RPA. In the MBPT2-RPA scheme the ground state wave-function
is constructed at the first order perturbation theory level and the EOM matrix is constructed in the one-hole (1h) or two-hole (2h) 
space for the ionization and double ionization potential respectively. MBPT2-EOMCC uses first order perturb 
wave function and the EOM matrix is constructed at 1h and two-hole and one-particle (2h-1p) space for the ionization 
problem whereas it is 2h and 3h-1p space for the double ionization problem.
The CCSD-RPA scheme uses the ground state wave-function at the coupled-cluster singles and doubles level and
EOM matrix is constructed at the 1h and 2h space for single ionization and double ionization respectively.
All these results are compared with the results calculated using CCSD-EOMCC method to understand passage 
of the electron correlation effects at various levels of approximation.

The manuscript is organized as follows: A brief discussion on the relativistic method to generate atomic
single particle orbital is presented in Sec. \ref{sec2}. This is followed by a description of the EOMCC theory 
in the context of evaluation of IPs and DIPs in Sec. \ref{sec3}. We present the results and discuss about 
their trends in Sec. \ref{sec4} before making our final remarks in Sec. \ref{sec5}. Atomic units (a.u.) is 
used consistently unless stated otherwise.

\section{Generation of basis and nuclear potential}\label{sec2}

We use the Dirac-Coulomb (DC) Hamiltonian in our calculations which, after scaling with the rest 
mass energy of the electrons ($c^2$), is given by
\begin{eqnarray}
H &=& \sum_i \left [ c\mbox{\boldmath$\alpha$}_i\cdot \textbf{p}_i+(\beta_i -1)c^2 + V_{nuc}(r_i) +
\sum_{j>i} \frac{1}{r_{ij}} \right ], \ \ \
\end{eqnarray}
where $\mbox{\boldmath$\alpha$}_i$ and $\beta_i$ are the Dirac matrices, $V_{nuc}(r_i)$ is the nuclear 
potential and $\frac{1}{r_{ij}}$ is the electron-electron repulsion potential.

The four-component Dirac wave function for an electron is given by
\begin{eqnarray}
 |\phi(r) \rangle = \frac {1}{r} \left ( \begin{matrix} P(r) &  \chi_{\kappa,m}(\theta,\phi)  \cr
                                                  iQ(r) & \chi_{-\kappa,m}(\theta,\phi) \cr
                                                        \end{matrix}  \right )
\end{eqnarray}
with $P(r)$ and $Q(r)$ are the large and small components of the wave function and $\chi_{\pm \kappa,m}(\theta,\phi)$
are the angular functions of the relativistic quantum number
$\kappa = -(j+\frac {1}{2})a$ satisfying the condition for the orbital angular momentum $l = j - \frac{a}{2}$ and
total angular momentum $j$. Linear combination of Gaussian type of orbitals (GTOs) is used to obtain the DF single 
particle orbitals $|\phi_{n,\kappa} (r) \rangle$ as
\begin{eqnarray}
 |\phi_{n,\kappa} (r) \rangle = \frac {1}{r} \sum_{\nu} \left (
         \begin{matrix}
         C_{n,\kappa}^L N_L f_{\nu}(r) & \chi_{\kappa,m} \cr
         i C_{n, -\kappa}^S N_S \left (\frac{1}{dr} + \frac{\kappa}{r} \right ) f_{\nu}(r) &\chi_{-\kappa,m}\cr
                 \end{matrix}
         \right ) \ \
\end{eqnarray}
where $n$ is the principal quantum number of the orbital, $C_{n,\kappa}$s are the expansion coefficients, 
$N_{L(S)}$ is the normalization constant for the large (small) component of the wave function and $\alpha_{\nu}$ is 
a suitably chosen parameter for orbitals of different angular momentum symmetries and $f_{\nu}(r) = 
r^l e^{-\alpha_{\nu} r^2}$ is a GTO. The even tempering condition $\alpha_{\nu} = \alpha_0 \beta^{\nu-1}$ with two 
parameters $\alpha_0$ and $\beta$ used for the exponents. The small component and large component of the wave 
function are related through the kinetic balance condition. The two parameter Fermi-charge distribution 
\begin{equation}
\rho_{nuc}(r)=\frac{\rho_{0}}{1+e^{(r-b)/a}},
\end{equation}
with the normalization factor $\rho_0$, the half-charge radius $b$ and $a= 2.3/4(ln3)$ is related to the skin 
thickness of the atomic nucleus, is considered for the evaluation of the nuclear potential.

After obtaining the single particle orbitals, we calculate the matrix element of the Coulomb interaction operator
using the expression
\begin{eqnarray}
\langle \phi_a \phi_b  | \frac {1}{r_{12}} | \phi_c \phi_d \rangle &=& \delta(m_a-m_c,m_d-m_b) \sum_{k,q} (-1)^{ m_a +m_b} \nonumber \\ &&\int dr_1 [P_a(r_1)P_c(r_1) + Q_a(r_1)Q_c(r_1)] 
 \nonumber \\ & \times & \int dr_2 [P_b(r_2)P_d(r_2) + Q_b(r_2)Q_d(r_2)] \nonumber \\
  & \times & \frac {r_<^k}{r_>^{k+1}}  \times X_k  \Pi(\kappa_a,\kappa_c,k) \Pi(\kappa_b, \kappa_d,k)
\end{eqnarray}
with the multipole $k$ determined by the triangular conditions $|j_a - j_c| \le k \le j_a + j_c$ and 
$|j_b - j_d| \le k \le j_b + j_d$ satisfying the condition for the function $\Pi(\kappa,\kappa',k) = 
\frac {1}{2} [1-aa'(-1)^{j+j'+k}]$ for $l + l' + k =$ even. The angular momentum factor $X_k$ is given by
\begin{eqnarray}
X_k &=& (-1)^{q+1} \sqrt{(2j_a+1)(2j_c+1)(2j_b+1)(2j_d+1)} \nonumber \\
 && \left ( \begin{matrix} j_a &  k & j_c  \cr
                            m_a & q & m_c \cr
                             \end{matrix}  \right )  \left ( \begin{matrix} j_b &  k & j_d  \cr
                            m_b & -q & m_d \cr
                             \end{matrix}  \right ) \nonumber \\ &&
   \left ( \begin{matrix} j_a &  k & j_c  \cr
                            1/2 & 0 & -1/2 \cr
                             \end{matrix}  \right )  \left ( \begin{matrix} j_b &  k & j_d  \cr
                            1/2 & 0 & -1/2 \cr
                             \end{matrix}  \right ).                         
\end{eqnarray}

\section{Brief description of the EOMCC theory}\label{sec3}

In the EOMCC approach, the wave function for the $k^{th}$ target state is created
by the action of a linear operator ($\Omega_k$) on the single reference CC wave function 
$|\Psi_{0}\rangle=\exp(\hat T)|\Phi_0\rangle$. i.e.  
\begin{eqnarray}
|\Psi_k\rangle &=& \Omega_k |\Psi_0 \rangle \nonumber \\ 
  &=& \Omega_k \exp(\hat T)|\Phi_0\rangle ,
\end{eqnarray}
where ${|\Phi_{0}\rangle}$ is a reference determinant which is taken as the Dirac-Fock (DF) wave function 
in the present case and the CC operator $\hat T$ accounts for the hole (h) to particle (p) excitations. 
The $T$ operators are given as
\begin{equation}
\hat T  = \sum\limits_{\stackrel{a < b \dots }{i < j \dots }}^N
t^{a b \dots }_{i j \dots} \ a^+ i b^+ j \dots ,
\end{equation}
in terms of strings of creation and annihilation operators for the holes (denoted by indices $i$, $j$, $\cdots$) and 
particles (denoted by indices ($a$, $b$, $\cdots$) for total $N$ number of electrons of a system.
We start with the energy eigenvalue equation 
\begin{equation}
\hat H \Omega_k \exp(\hat T)|\Phi_{0}\rangle=E_{k} \Omega_k \exp(\hat T)|\Phi_{0}\rangle.
\end{equation}
By operating on both the sides of the equations with $exp(-\hat T)$ from the left-hand side and considering
$\Omega_k$ and $\hat T$ commute each other owing to the fact that they are made up of strings of same 
quasi-particle creation and annihilation operators, we get 
\begin{equation}
\bar{H} \Omega_k |\Phi_{0}\rangle=E_{k} \Omega_k|\Phi_{0}\rangle,
\end{equation}
for ${\bar {H}\equiv exp(-\hat T)\,\hat H\,exp(\hat T)}$. The above equation is projected onto the basis of 
excited determinants those are accessible by the action of $\Omega _k$ on $|\Phi_{0}\rangle$. To simplify the
notations, we denote the IP evaluating EOMCC operators as 
\begin{equation}
\begin{split}
\Omega_k^{\text{IP}}  \equiv & R_1+R_2+ \cdots\\
             =&  \sum_{i} r_{i} i +  \sum\limits_{\stackrel{a}{i < j}} r_{i j}^{a} a^{+} j i + \cdots ,
\end{split}
\end{equation}
whereas for DIP evaluation, we define  
\begin{equation}
\begin{split}
\Omega^{\text{DIP}}_k \equiv & S_2 +S_3 + \cdots\\
             =&  \sum_{i<j} s_{i j} j i +  \sum\limits_{\stackrel{a}{i < j < k}} s_{i j k}^{a} a^{+} k j i + \cdots .
\end{split}
\end{equation}
We project on to the set of excited determinants ($|\Phi_i\rangle$) and ($|\Phi_{ij}^a\rangle$) representing 
the 1h and 2h-1p determinants, respectively, for the evaluation of the $R_1$ and $R_2$ amplitudes as
\begin{equation}
{\langle \Phi_{i}|\bar{H} R_k|\Phi_{0}\rangle=E_k \langle \Phi_{i}|R_k|\Phi_{0}\rangle}
\end{equation}
and
\begin{equation}
{\langle \Phi_{ij}^{a}|\bar{H} R_k|\Phi_{0} \rangle =E_k \langle \Phi_{ij}^{a}|R_k|\Phi_{0}\rangle}.
\end{equation}
Similarly, we project $|\Phi_{ij}\rangle$ and $|\Phi_{ijk}^{a}\rangle$ representing 2h and 3h-1p determinants, 
respectively, to determine the $S_1$ and $S_2$ amplitudes as
\begin{equation}
{\langle \Phi_{ij}|\bar{H} S_k |\Phi_{0}\rangle=E_k \langle \Phi_{ij}|S_k|\Phi_{0}\rangle},
\end{equation}
and
\begin{equation}
{\langle \Phi_{ijk}^{a}|\bar{H} S_k|\Phi_{0} \rangle =E_k \langle \Phi_{ijk}^{a}|S_k|\Phi_{0}\rangle}.
\end{equation}
It has to be noted that the 3h-1p excitations for the He-like systems are absent, hence only the 2h 
projections are made in this case after performing the CC calculations.
All these equations are expressed in the matrix form as
\begin{equation}
\bar{H}_N \Omega_k = \Omega_k \Delta E_k
\end{equation}
to solve for the eigenvalues $\Delta E_k=E_k-E_0$, the energy difference between the $|\Psi_0\rangle$ state $E_0$ and 
the ionized $|\Psi_k\rangle$ state $E_k$, and their corresponding eigenvectors by using the normal order Hamiltonian
($H_N$). The Davidson algorithm \cite{72}, which is an iterative diagonalization scheme, is implemented to diagonalize 
the non-hermitian matrix elements of the effective Hamiltonian ${\bar H}_N$. We have constructed the EOMCC matrix 
in the opted space but solve only for the principal peaks.
\begin{figure}
\begin{center}
\includegraphics[width=8.0cm, height=8.0cm]{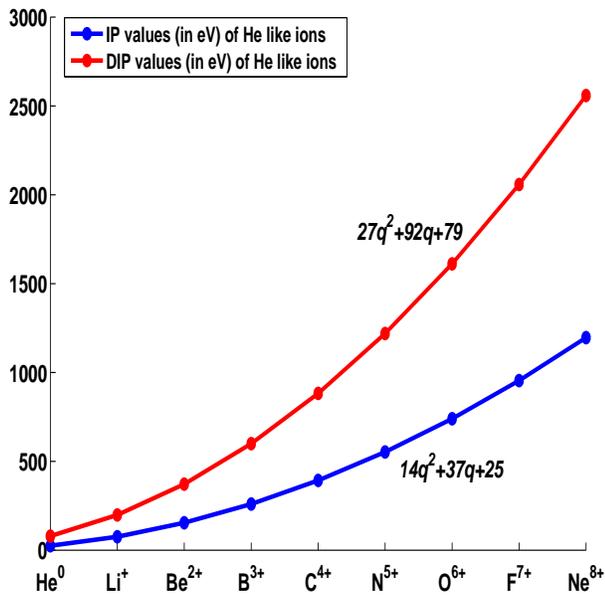}
\end{center}
\caption{Trends of IPs and DIPs in the He-like systems as a function of their ionic charge ($q$).}
\label{fig2}
\end{figure}

\section{Results and discussions}\label{sec4}

Here, we present and discuss about the numerical results of the calculated IPs and DIPs of the He-like and 
Be-like systems considering atoms up to Ne using the relativistic EOMCC methods at different levels of approximations. 
To carry out these calculations, we construct the single particle orbitals in the universal basis (UB) method using GTOs with 
$\alpha_0=0.004$ and $\beta=2.23$ for all the considered systems. For each of the atomic system, we use 40, 39, 38,
37 and 36 number of GTOs for the $s$, $p$, $d$, $f$ and $g$ symmetry waves in the DF method to obtain the self-consistent
field (SCF) solutions for the orbitals. For the EOMCC calculations, we only correlate electrons belonging to the 
low-lying orbitals up to $16s$, $14p$, $13d$, $11f$ and $10g$ as high-lying orbitals contribute less in the calculations of IPs and DIPs owing to their high energy values. In 
Table \ref{tab1}, we present the SCF ($ E^0_{DF}$) and correlation energies calculated at the the second order 
perturbation theory approximation ($ E^2_{corr}$) and from the CC method considering only the singles and doubles 
excitations, which is referred to as CCSD method in the literature, ($ E^{ccsd}_{corr}$).

\begin{table}[h]
\caption{SCF energy (${E_{DF}^{0}}$), correlation energies from the MBPT(2) (${E_{corr}^{(2)}}$) and
CCSD (${E_{corr}^{(ccsd)}}$) methods for different systems obtained using our calculations.}
\begin{ruledtabular}
\begin{tabular}{lccc}
System  & {${E_{DF}^{0}}$} & 
{${E_{corr}^{(2)}}$}& {${E_{corr}^{(ccsd)}}$}  \\
\hline
He          &\,   $-$2.8618  &\,  $-$0.0368  &\, $-$0.0417\\
Li$^{+}$    &\,   $-$7.2372  &\,  $-$0.0394  &\, $-$0.0429\\
Be$^{2+}$   &\,  $-$13.6139  &\,  $-$0.0406  &\, $-$0.0434\\
B$^{3+}$    &\,  $-$21.9931  &\,  $-$0.0412  &\, $-$0.0435\\
C$^{4+}$    &\,  $-$32.3759  &\,  $-$0.0415  &\, $-$0.0435\\
N$^{5+}$    &\,  $-$44.7641  &\,  $-$0.0416  &\, $-$0.0434\\
O$^{6+}$    &\,  $-$59.1597  &\,  $-$0.0416  &\, $-$0.0432\\
F$^{7+}$    &\,  $-$75.5650  &\,  $-$0.0415  &\, $-$0.0429\\                       
Ne$^{8+}$   &\,  $-$93.9827  &\,  $-$0.0412  &\, $-$0.0426\\
Be          &\,  $-$14.5758  &\,  $-$0.0745  &\, $-$0.0927\\ 
B$^{+}$     &\,  $-$24.2451  &\,  $-$0.0856  &\, $-$0.1091\\
C$^{2+}$    &\,  $-$36.4251  &\,  $-$0.0948  &\, $-$0.1236\\
N$^{3+}$    &\,  $-$51.1114  &\,  $-$0.1029  &\, $-$0.1371\\
O$^{4+}$    &\,  $-$68.3143  &\,  $-$0.1104  &\, $-$0.1500\\
F$^{5+}$    &\,  $-$88.0271  &\,  $-$0.1173  &\, $-$0.1623\\
Ne$^{6+}$   &\,  $-$110.2559 &\,  $-$0.1238  &\, $-$0.1742\\
\end{tabular}
\end{ruledtabular}
\label{tab1}
\end{table}

\begin{table*}[ht]
\begin{ruledtabular}
\caption{Comparison of our calculated IPs of He-like ions (in eV) with the NIST database. Differences are given as $\delta$
in \%.}
\begin{tabular}{l r r r  r r  r}
 System & MBPT2-RPA & MBPT2-EOMCC  &  CCSD-RPA  &   CCSD-EOMCC   &   NIST \cite{71}  &   $\delta$ \\
\hline
He      &\,25.9828  &\,24.4648     &\,26.1127   &\, 24.5850      &\,24.5873          &\,0.009   \\
Li$^+$  &\,77.0651  &\,75.5492     &\,77.1575   &\,75.6379       &\,75.6400          &\,0.002   \\
Be$^{2+}$  &\,155.3441 &\, 153.8286   &\, 155.4175 &\,153.9001      &\,153.8961         &\,0.002   \\
B$^{3+}$   &\,260.8492 &\, 259.3337   &\, 260.9107 &\,259.3941      &\,259.3715         &\,0.008  \\
C$^{4+}$   &\,393.6042 &\,392.0884    &\,393.6575  &\,392.1409      &\,392.0905         &\,0.012  \\
N$^{5+}$   &\,553.6337 &\,552.1174    &\,553.6810  &\,552.1641      &\,552.0673          &\.0.017  \\
O$^{6+}$   &\,740.9648 &\,739.4476    &\,741.0076  &\,739.4899      &\,739.3267          &\,0.022  \\
F$^{7+}$   &\,955.6285 &\,954.1102    &\,955.6677   &\,954.1491      &\,953.8980          &\,0.026  \\
Ne$^{8+}$  &\,1197.6602 &\,1196.1406   &\,1197.6965 &\,1196.1766    &\,1195.8077         &\,0.030  \\
\end{tabular}
\label{tab2}
\end{ruledtabular} 
\end{table*}
In Table \ref{tab2}, we present IPs of the He-like systems starting from He to Ne. In the same table, we also compare our 
results with the values listed in the NIST database \cite{71} and the deviations of our results from the NIST values are 
given in percentage as $\delta$.
Three intermediate calculations are also done for the He like ions. In MBPT2-RPA scheme both
dynamic and non-dynamic parts of the correlations are missing in comparison to the CCSD-EOMCC method. It is observed 
that the MBPT2-RPA values are overestimated than CCSD-EOMCC. It is also true for the scheme CCSD-RPA. Missing non-dynamic
correlations leads to these overestimations. The MBPT2-EOMCC scheme underestimates the results but these 
values are in better agreement than the other two scheme in comparison to the experimental values from NIST. 
It is observed from this table that with increase in the atomic charge, the deviation increases except for 
He for which it comparatively shows large discrepancy. We attribute the reason for the same as orbitals of the ions are more 
contracted towards the nucleus than He. Again, increase in discrepancies in ions with higher Z values indicates that
contributions from the neglected Breit interaction and quantum electrodynamic (QED) corrections are important to be
considered to improve accuracies in these results. Similarly, we present DIPs of the He-like ions in Table 
\ref {tab3} and compare them with the data available in the NIST database. Also, results from our calculations 
using MBPT2-DIP scheme are presented where the ground state wave function is the first order perturb wave 
function. In DIP calculations of He like systems the 3h-1p block does not contribute naturally.
Deviations of our DIP results of these systems from NIST are given as $\delta$ in percentage in the same table. This comparison between
the MBPT2-DIP and CCSD-DIP results shows that MBPT2-DIP values are underestimated than the CCSD-DIP values. 
This study demonstrates that though MBPT2-DIP method constructs the ground state wave function at the first 
order perturbation level but it gives reasonably accurate results by saving enormous computational time. 

\begin{table}[ht]
\caption{Comparison of DIPs of He-like ions (in eV) from our calculations and IERM method with NIST database. 
Differences between our values with NIST data are given as $\delta$ in \%.}
\begin{ruledtabular}
\begin{tabular}{lrrrrr}
System    &    MBPT2-DIP    &   CCSD-DIP    &   NIST \cite{71}     &  $\delta$  \\
\hline
He        &\, 78.8769       &\,79.0141      &\, 79.0051             &\,0.011 \\
Li$^{+}$  &\, 198.0081      &\,198.1053     &\,198.0935             &\,0.005  \\
Be$^{2+}$ &\, 371.5618      &\,371.6384     &\, 371.6152            &\,0.006   \\
B$^{3+}$  &\,599.5864       &\,599.6502     &\,599.5977             &\,0.008   \\
C$^{4+}$  &\,882.1263       &\,882.1813     &\,882.0847             &\,0.010    \\
N$^{5+}$  &\,1219.2303      &\,1219.2788    &\,1219.1142            &\,0.013    \\
O$^{6+}$  &\,1610.9541      &\,1610.9977    &\,1610.7380            &\,0.016    \\
F$^{7+}$  &\,2057.3616      &\,2057.4013    &\,2057.0158            &\,0.018    \\
O$^{8+}$  &\,2558.5256      &\,2558.5621    &\,2558.0076            &\,0.021    \\
\end{tabular}
\label{tab3}
\end{ruledtabular}
\end{table}

\begin{figure}[ht]
\centering
\begin{center}
\includegraphics[width=8.0cm, height=8.0cm]{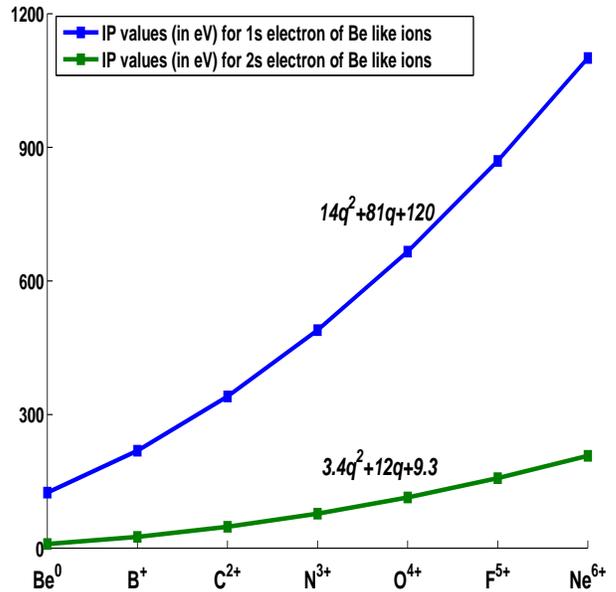}
\end{center}
\caption{Trends of IP values in the Be-like systems as a function of ionic charge ($q$).}
\label{fig4}
\end{figure}
In order to assess the trends followed by the calculated results using different employed methods with the ionic charge ($q$) of the considered atomic systems, we 
plot the calculated IPs and DIPs of the He-like systems as a function of $q$ as shown in Fig. \ref{fig2}. We found that both the 
IPs and DIPs follow the standard quadratic equations $aq^2+bq+c$ with arbitrary parameters $a$, $b$ and $c$. It is
found that DIPs of He-like ions satisfy the relationship $27q^2+92q+79$ while for the IPs obey $14q^2+37q+25$ 
trend. The reason for large $a$ coefficient for DIPs may be owing to large kinetic energies of the 
electrons in the doubly ionized systems than the singly ionized systems. Presence of linear terms in $q$ with different 
magnitudes of $b$ coefficient for IPs and DIPs correspond to the role of the Coulomb interactions. The larger
ratios of $b/a$ and $c/b$ in DIPs imply dominant role by the kinetic energies than the Coulomb interactions in the 
evaluation of DIPs in the heavier ions.
\begin{figure}
\centering
\begin{center}
\includegraphics[width=8.0cm,height=8.0cm]{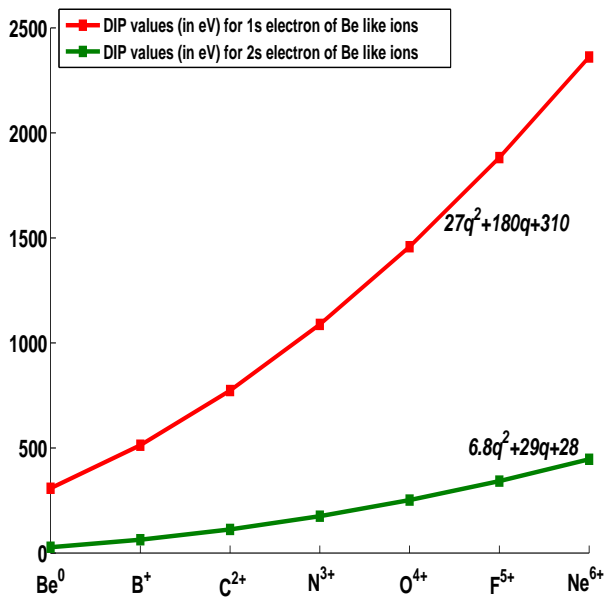}
\end{center}
\caption{Trends of DIP values in the Be-like systems as a function of ionic charge ($q$).}
\label{fig6}
\end{figure}
\begin{table*}
\begin{ruledtabular}
\caption{Comparison of our calculated IPs of Be-like ions (in eV) with the NIST database. Differences are given as $\delta$
in \%.}
\begin{tabular}{l r r r r r r r  r r  r}
System & \multicolumn{2}{c}{MBPT2-RPA} & \multicolumn{2}{c}{MBPT2-EOMCC}  &\multicolumn{2}{c}{CCSD-RPA}  & \multicolumn{2}{c}{CCSD-EOMCC}&NIST \cite{71}  &   $\delta$ \\
\cline{2-3} \cline{4-5}  \cline{6-7} \cline{8-9}\\
    & $1s$  & $2s$  &$1s$  &$2s$ & $1s$  & $2s$  & $1s$  & $2s$ &$2s$  \\
\hline    
Be  &\,129.9469 &\, 9.3047&\,124.7181&\,8.9465&\,129.7176&\,9.6607&\,124.6451&\,9.3252&\,9.3226&\,0.028\\
B$^{+}$&\,223.9840&\,24.9441&\,218.8594&\,24.6074&\,223.7854&\,25.4713&\,218.7644&\,25.1534&\,25.1548&\,0.005\\
C$^{2+}$&\,345.5711&\,47.5097&\,340.6506&\,47.1797&\,345.3836&\,48.1981&\,340.5585&\,47.8859&\,47.8877&\,0.003\\
N$^{3+}$&\,494.5523&\,76.9378&\,489.5203&\,76.6103&\,494.3698&\,77.7834&\,489.3986&\,77.4733&\,77.4735&\,$\sim$0.000\\
O$^{4+}$&\,670.8905&\,113.2110&\,665.8323&\,112.8843&\,670.7105&\,114.2119&\,665.6995&\,113.9024&\,113.8989&\,0.003\\
F$^{5+}$&\,874.5887&\,156.3277&\,869.6562&\,156.0011&\,874.4103&\,157.4820&\,869.5240&\,157.1727&\,157.1631&\,0.006\\
Ne$^{6+}$&\,1105.6664&\,206.2915&\,1100.7049&\,205.9644&\,1105.4891&\,207.5978&\,1100.5646&\,207.2880&\,207.2710&\,0.008\\
\end{tabular}
\label{tab4}
\end{ruledtabular}
\end{table*}

In Table \ref{tab4}, we tabulate the IP values for both the $1s$ and $2s$ orbitals of the considered Be-like ions. 
The results of the $2s$ orbitals are compared with the NIST values and the deviations, given in percentage as $\delta$
from the NIST values, are found to be very small. The DIP values are given in Table \ref{tab5} and are compared with the 
NIST data. As can be seen, we have achieved accuracies of less than 
$\sim 0.01\%$ for all the Be-like systems. For Ne$^{6+}$, the absolute value differs by 0.04 eV and for the rest it is 
only about 0.01 eV. For Be like systems we also present intermediate results using MBPT2-RPA, MBPT2-EOMCC and CCSD-RPA
schemes. It has been observed that the RPA values are overestimated than the CCSD-EOMCC results. On the other hand MBPT2-EOMCC
values are lower than the CCSD-EOMCC values. The differences between the results from these two schemes are 
more for the 2s orbitals where as less for the 1s orbitals.

\begin{table*}[ht]
\begin{ruledtabular}
\caption{Comparison of our calculated DIPs of Be-like ions (in eV) with the NIST database. Differences are given as $\delta$
in \%.}
\begin{tabular}{l r r r r r r r  r r  r}
System & \multicolumn{2}{c}{MBPT2-RPA} & \multicolumn{2}{c}{MBPT2-EOMCC}  &\multicolumn{2}{c}{CCSD-RPA}  & \multicolumn{2}{c}{CCSD-EOMCC}&NIST \cite{71}  &   $\delta$ \\
\cline{2-3} \cline{4-5}  \cline{6-7} \cline{8-9}\\
    & $1s$  & $2s$  &$1s$  &$2s$ & $1s$  & $2s$  & $1s$  & $2s$ &$2s$  \\
\hline    
Be  &\,321.0077 &\,27.1344&\,308.5294&\,27.0969&\,320.4761&\,27.5600&\,308.3921&\,27.5427&\,27.5338&\,0.032\\
B$^{+}$&\,526.2121&\,62.5409&\,513.4659&\,62.4882&\,525.7490&\,63.1252&\,513.2750&\,63.0833&\,63.0869&\,0.005\\
C$^{2+}$&\,786.4974&\,111.6955&\,773.6588&\,111.6329&\,786.0613&\,112.4343&\,773.4375&\,112.3789&\,112.3803&\,0.001\\
N$^{3+}$&\,1101.5640&\,174.5374&\,1088.4372&\,174.4678&\,1101.1411&\,175.4290&\,1088.1864&\,175.3649&\,175.3638&\,$\sim$0.000\\
O$^{4+}$&\,1471.3467&\,251.0545&\,1458.4205&\,250.9797&\,1470.9312&\,252.0981&\,1458.2743&\,252.0278&\,252.0183&\,0.003\\
F$^{5+}$&\,1895.8582&\,341.2520&\,1882.5409&\,341.1733&\,1895.4476&\,342.4468&\,1882.2607&\,342.3718&\,342.3493&\,0.006\\
Ne$^{6+}$&\,2375.1422&\,445.1430&\,2361.6537&\,445.0611&\,2374.7351&\,446.4879&\,2361.3575&\,446.4093&\,446.3676&\,0.009\\
\end{tabular}
\label{tab5}
\end{ruledtabular}
\end{table*}
In Fig. \ref{fig4}, we plot the IP values of the inner $1s$ electron of Be-like ions against their corresponding $q$ values and find also
a parabolic behavior with $a$ coefficient having same value as of the plot for the IPs of the He-like ions. However, since the Coulomb 
potentials are larger in the Be-like systems for which its $b$ coefficient found to be larger. In the same figure, we also give 
plot for IPs of the outer $2s$ electrons from the respective Be-like systems and find again a parabolic trend but with
much smaller $a$, $b$ and $c$ coefficients. This may be due to the the fact that the outer $2s$ electrons are loosely bound with
their nucleus. Similarly, we plot the DIPs of Be-like systems for both the $1s$ and $2s$ orbital electrons in Fig. 
\ref{fig6}. We see like-wise plots with Fig. \ref{fig4} and find for the $1s$ orbital electrons it follows the relationship 
$27q^2+180q+310$ while for the $2s$ orbital electrons it obeys $6.8q^2+29q+28$ relation with their ionic charges. These
coefficients are almost twice than that of their corresponding IPs. Again, when we compare trends of DIPs of the $1s$
orbital electrons of Be-like ions with the He-like ions, we find the corresponding $a$ coefficient is same but the $b$
coefficient value is large.

\section{Conclusion}\label{sec5}

We have developed and applied four-component EOM methods at different levels of approximations to evaluate IPs 
and DIPs of the He-like and Be-like isoelectronic sequences up to Ne. Comparison between our results with the 
NIST database shows that EOMCC results are in very good agreement with the values tabulated in NIST database. It has been
demonstrated that both IPs and DIPs of He-like and Be-like isoelectronic sequences follow parabolic trends and the
magnitudes of the coefficients of the parabolas are attributed to the role of the kinetic energies and Coulomb
interactions of the systems. These coefficients for the $1s$ orbital electrons are almost twice compared to the
$2s$ orbital electrons in the Be-like systems. Excellent agreement of our calculated results with the NIST values 
implies that our EOMCC method is capable of accounting for both the relativistic and electron correlation effects
accurately. Observation of slight deviations in the values for heavier ions imply that inclusion of higher 
order relativistic corrections may be required to improve accuracies of those results.

\section*{Acknowledgment}
H.P., T.S., N.V. and S.P. acknowledge a grant from CSIR XIIth Five Year Plan project on Multi-Scale Simulations of 
Material (MSM) and the resources of the Center of Excellence in Scientific Computing at CSIR-NCL. H.P. acknowledge the 
Council of Scientific and Industrial Research (CSIR) for fellowship. S.P. acknowledges the award of the 
J. C. Bose Fellowship from DST. B.K.S acknowledges 3TFlop HPC cluster at Physical Research Laboratory for part of the
calculations.

\end{document}